\documentclass[10pt,twoside,twocolumn]{IEEEtran}
\usepackage{graphicx,cite,calc,epstopdf}
\usepackage{tikz}
\usetikzlibrary{shapes,arrows}
\usepackage{rotating}
\usepackage{amsmath}
\ifCLASSINFOpdf
\else
\fi

\begin{document}
%
\title{An Optimal Detector for Hierarchical $2^n$ QAM }


\author{Hadi~Alasti,~\IEEEmembership{Senior Member,~IEEE}

\thanks{H. Alasti is with the Department of Computer, Electrical and Information Technology, Indiana University-Purdue University, Fort Wayne (IPFW), Fort Wayne, IN, 46805 USA (e-mail: halasti@ieee.org).}
}


%


\maketitle

\begin{abstract}
A practical and scalable, optimal detection algorithm is introduced for hierarchical Gray-coded $2^n$ QAM for application in software defined communications. The introduced optimal detection algorithm is based on successive interference cancellation and is equivalent to a successive approximation analog to digital conversion (SAR-ADC). The performance of the optimal detector is discussed in additive white Gaussian noise channels. The proposed detection algorithm needs much less storage space than the conventional optimal detection algorithm followed by table lookup. 
\end{abstract}

\IEEEpeerreviewmaketitle

\section{Introduction}
Hierarchical modulations with their performance-selective attribute, provide different data services to the users based on their reception condition from the channel. This modulation is one of the possible approaches to provide variable rates at receiver when proper codecs such as IEEE H.264 is available~\cite{Schwarz,ETSI}. 
Besides the good spectral efficiency of quadrature amplitude modulations (QAMs), hierarchical $2^n$ QAM has simpler implementation with less complexity \cite{vectored_QAM}. QAM modulations with large constellations are used in or proposed for various modern communication systems such as fast wireless local area networks under IEEE 802.11ax standards \cite{IEEE802.11ax}, long term evolution (LTE) under IEEE 802.16x \cite{IEEE802.16x}, fast optical communications~\cite{Fast-optical}, Ethernet protocol over coax \cite{Gorshe}, etc.

In $2^n$ QAM, each $n$ bits of binary information, or so-called one code-word are multiplexed on analog voltage levels pair of in-phase  and quadrature or in short (\emph{I}, \emph{Q}). In radio frequency section, \emph{I} and \emph{Q} modulate a given single frequency $f_0$ at 90 degrees phase shift, which makes them orthogonal. At receiver, after compensation of the channel's destructive effects, an estimate of \emph{I} and \emph{Q} is given to the detector to decide about a most likely sent binary code-word ~\cite{Gallager}. In this paper, the baseband (\emph{I}, \emph{Q}) pair is interchangeably called \emph{I} and \emph{Q} branches.

\begin{figure}[t!]
	\centering
		\includegraphics[width=3.3in,angle=0]{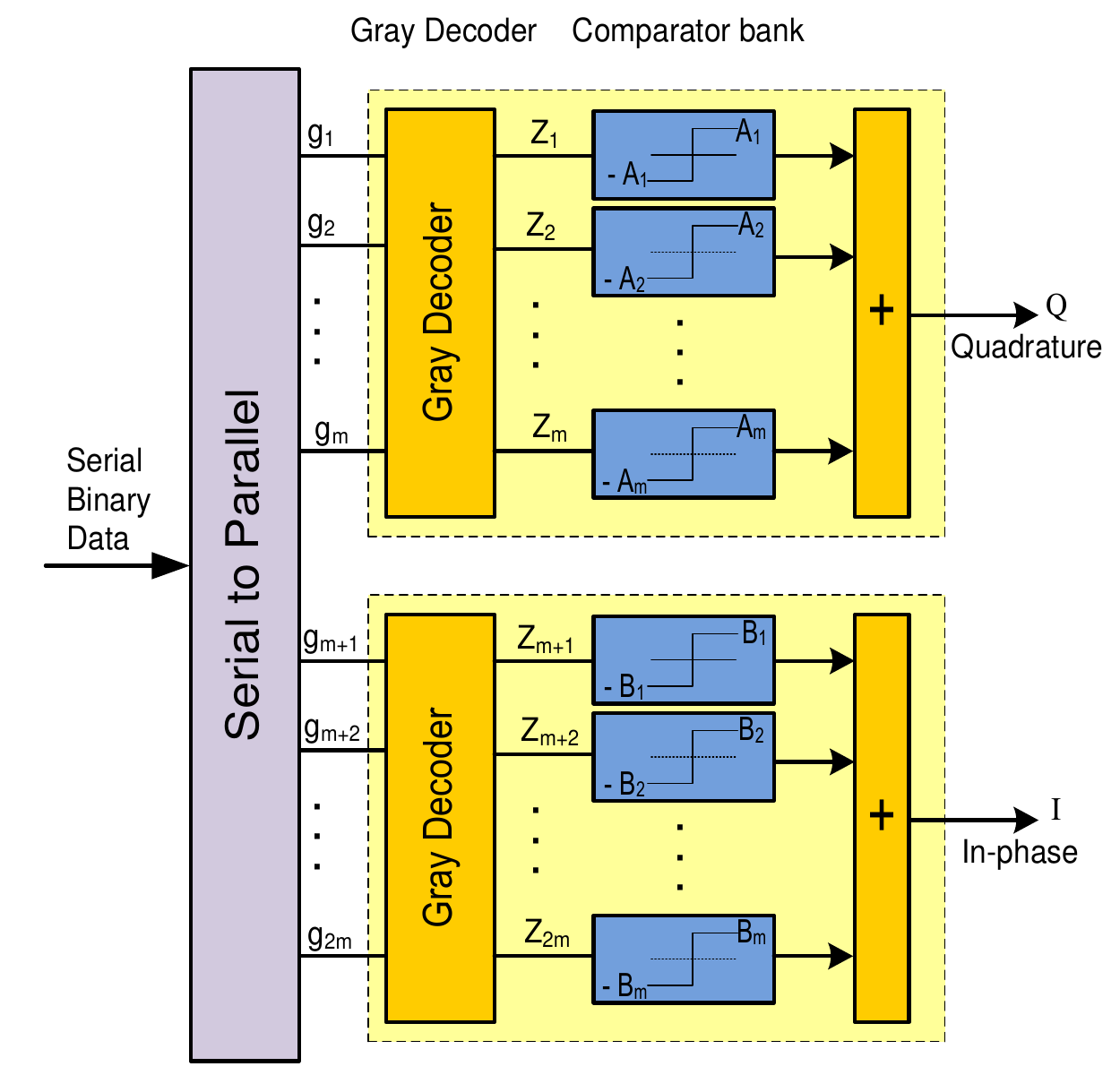}
		\caption{Gray-coded hierarchical $2^{2m}-$QAM modulator.}\label{fig: 1}
\end{figure}

The layered structure of hierarchical QAM lets reduce the implementation complexity of the transceiver and flexibly allocate the information rate \cite{Hua-Sun}. Vectored implementation of $2^{2m}$ QAM was introduced in \cite{vectored_QAM} and generalized in \cite{Hadi-CCNC} as a complexity reduced approach at transmitter and receiver. In the present paper, we show that the introduced detection algorithm in \cite{vectored_QAM} for receiver has the same performance as maximum likelihood (ML) optimum detection algorithm with hard decision decoder. The specific constellation bit mapping introduced in \cite{vectored_QAM} splits the Gray decoder at transmitter and receiver, where it completely isolates the multiplexed information on \emph{I} and \emph{Q} branches. Fig.~\ref{fig: 1}, shows the complexity reduced vectored implementation of $2^{2m}$ QAM modulator. This implementation models the $2^{2m}$ QAM as two groups of co-existing binary phase shift keying (BPSK) sub-channels in \emph{I} and \emph{Q} baseband branches, each with $m$ interfering sub-channels. The BPSK sub-channel $\# i$  in \emph{Q} branch is projected to a known level of $\pm A_i$ in a comparator. Similarly, each BPSK sub-channel $\#(i+m)$ in \emph{I} branch is projected to a known level of $\pm B_i$, as it is illustrated in Fig.~\ref{fig: 1}. In this figure, $g_i$ and $z_i$ are 0 or 1.  Each sub-channel carries a single bit of modulation's code-word. Finally, these gained sub-channels are added together to form the \emph{I} and \emph{Q} outputs. The simplicity of adjusting the amplification gain of the sub-channels make this model a tractable implementation model for performance adaptation in software defined communication systems such as software defined radios.

This paper is organized as follows: in the next section vectored implementation of $2^{n}$ QAM is reviewed. In section III, the optimum detection algorithm for $2^n$ QAM is discussed. Then, in section IV, the performance of the detection algorithm is presented.
\section{Vectored implementation of $2^n$ QAM}
Vectored implementation of hierarchical $2^{2m}$ QAM was introduced in \cite{vectored_QAM} as a practical, scalable and complexity reduced approach. To generate $2^{2m-1}$-QAM modulator, let us eliminate the sub-channel $\# 2m$ in \emph{I} branch and drop its related most significant bit (MSB). By dropping the MSB of the zig-zag code-word, the half left and right sides of the constellation (see Fig. 2-(a)) shift towards the origin until they coincide. This transform turns the constellation of Fig. 2-(a) into Fig. 2-(b). The MSB drop turns the $2m$-bit Gray code into $2m-1$-bit Gray code. As such, Gray-coded $2^{2m-1}$ QAM modulator is made. As such, continuing this process and eliminating the $m-k$ MSB sub-channels from \emph{I} branch, $2^{k+m}$ QAM modulator is generated.\\
\begin{figure}[t]
  \centering
  \begin{tabular}{p{.52\columnwidth}p{.32\columnwidth}}
    \includegraphics[width=1.55in,angle=0]{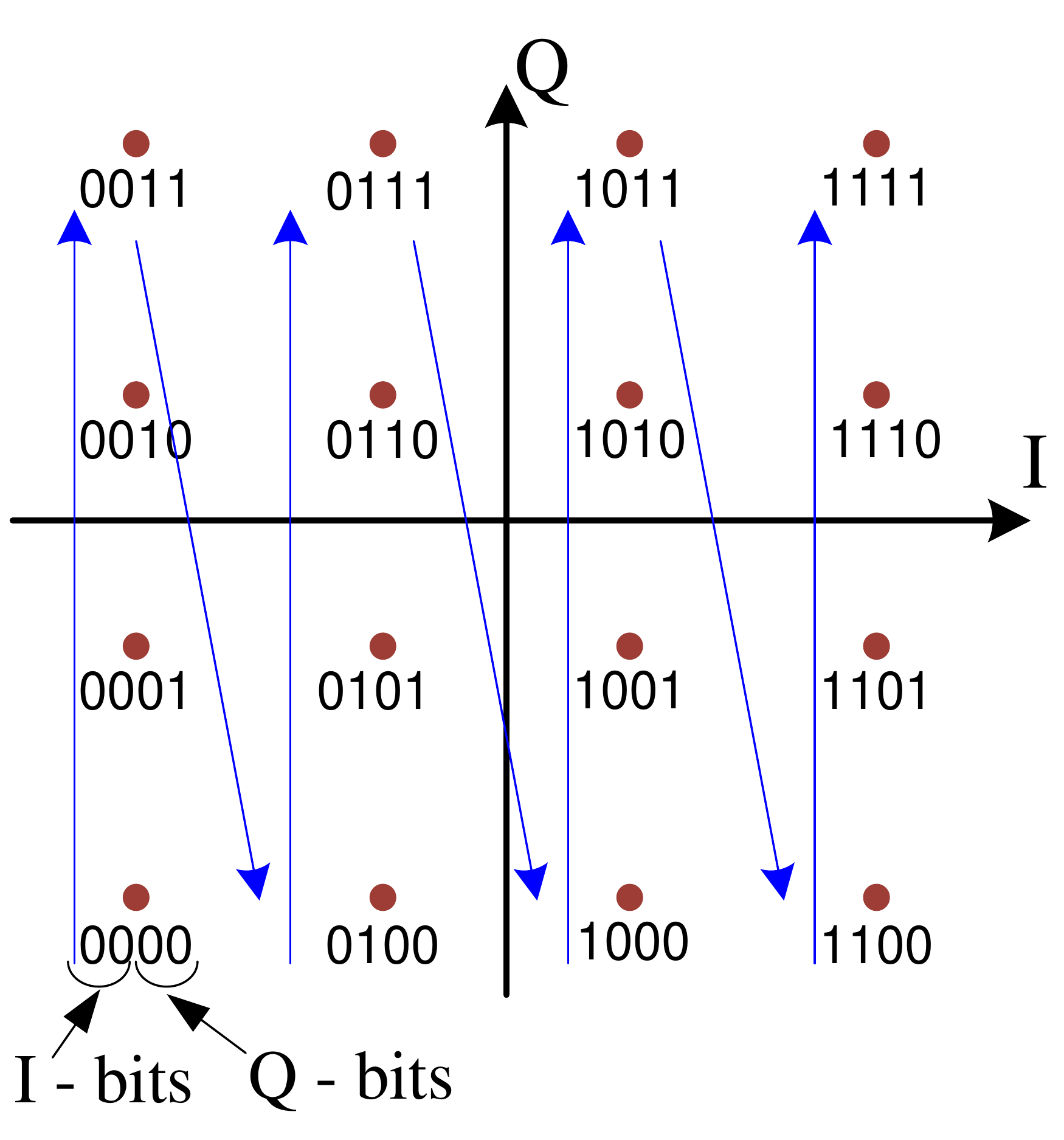}&
    \includegraphics[width=1.1in,angle=0]{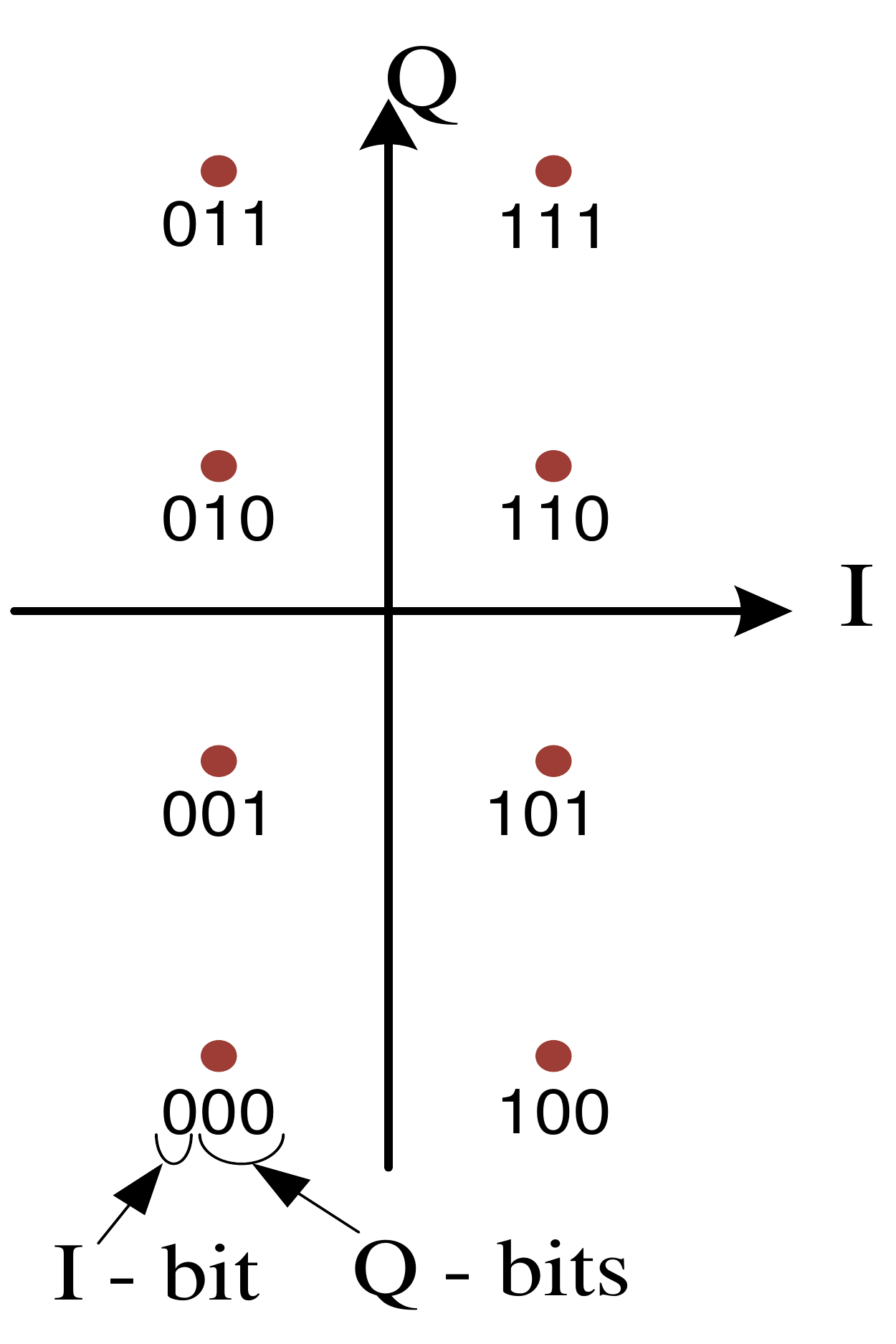}\\
    \footnotesize (a)  Zig-zag bit mapping for 16-QAM. & \footnotesize (b) Formation of 8-QAM from 16-QAM.\\
  \end{tabular}  
  \label{fig: 2}\caption{Zig-zag constellation bit-mapping for 16-QAM and formation of 8-QAM by dropping the zig-zag code-word's MSB in 16-QAM (before Gray-coding).}
\end{figure}
\begin{figure}
		\centering
		\includegraphics[width=1.8in, angle=0]{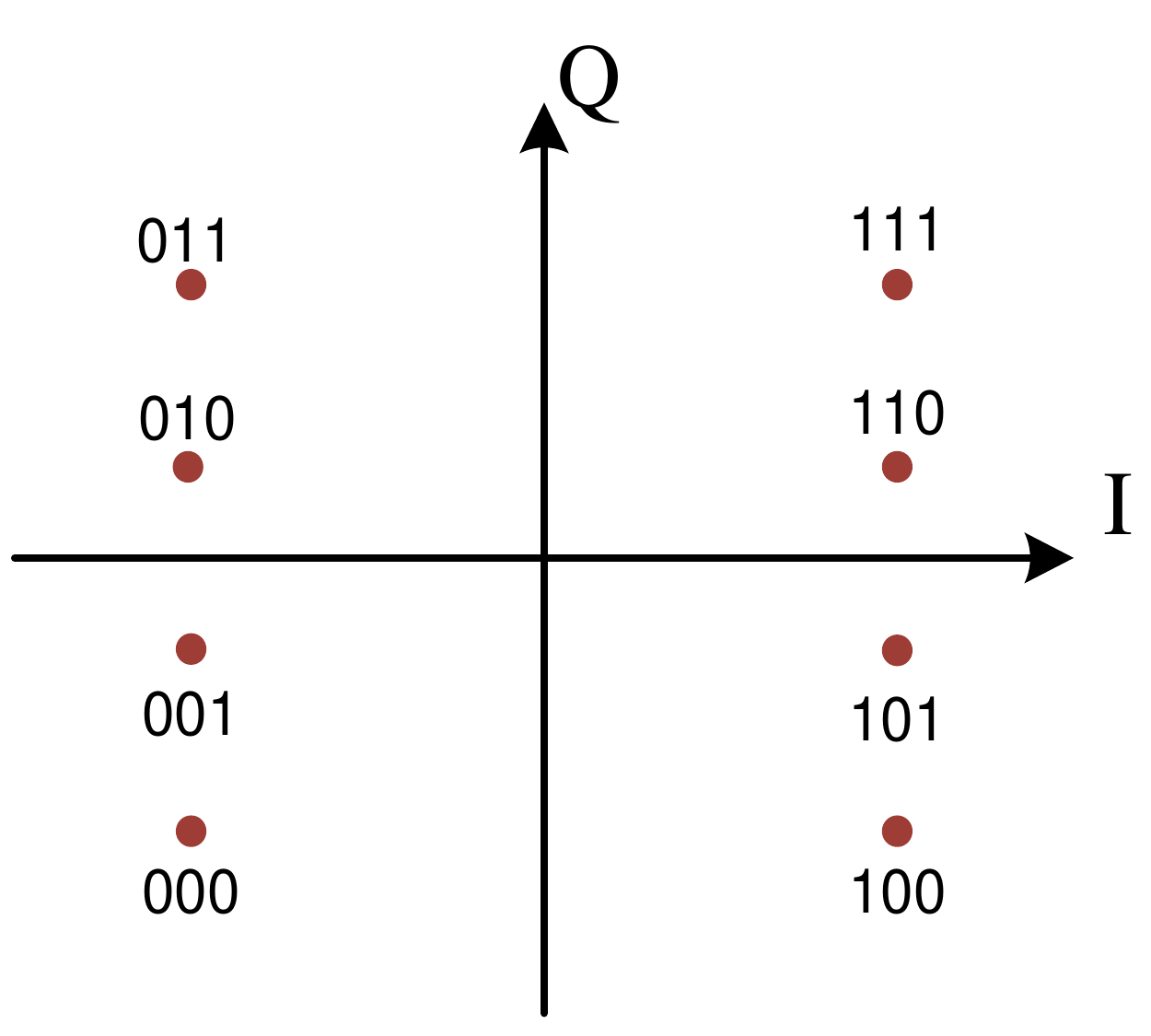}
		\caption{Stretching the 8-QAM constellation to reduce PAPR.}
		\label{fig: 3}
\end{figure}
Equal space between the symbols in  \emph{I} and \emph{Q} directions in $2^{2m-1}$ QAM (see Fig. 2-(b) for 8-QAM), results in higher peak-to-average-power-ratio (PAPR) than $2^{2m}$ QAM, which is not desirable. To reduce this PAPR, constellation can be stretched, where it significantly improves the bit error rate (BER) of the \emph{I} branch's bits, as well. Fig.~\ref{fig: 3} shows the constellation of 8-QAM after stretch towards \emph{I}.

\begin{figure}[b]
	\centering
		\includegraphics[width=2.5in,angle=0]{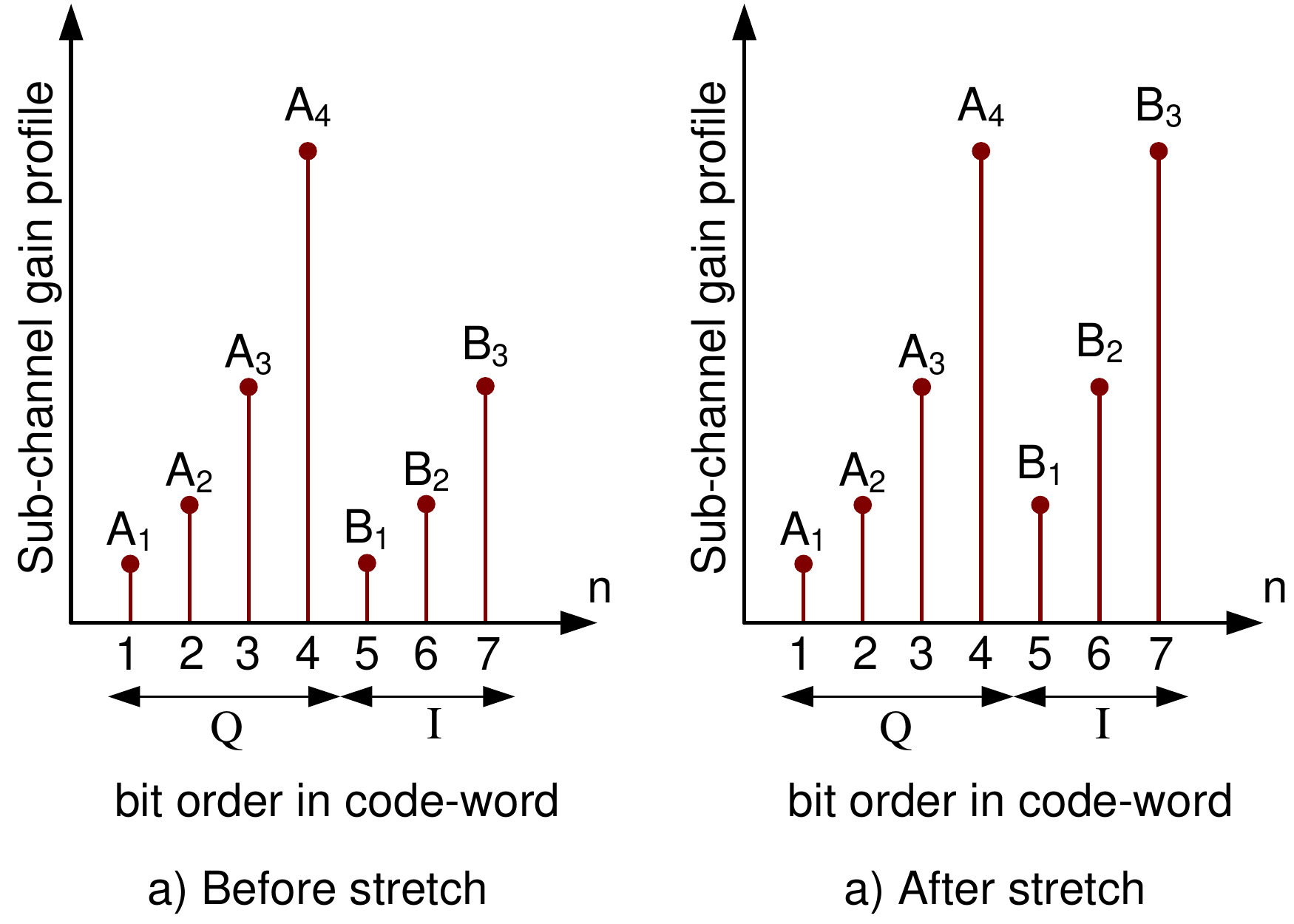}
		\caption{sub-channel gain profile of 128-QAM.}\label{fig: 4}
\end{figure}

For implementation of $2^{k+m}$ QAM, we use the model of $2^{2m}$ QAM modulator, which is illustrated in Fig.~\ref{fig: 1}. This model is completely introduced by two sets of positive sub-channel gains of $\{B_j\}_{j=1}^k$ and $\{A_j\}_{j=1}^m$. According to this model, the \emph{I} and \emph{Q} branches of QAM are formed by addition of multiple co-existing BPSK sub-channels, where each sub-channel's gain is equal to one of the defined values in $\{B_j\}_{j=1}^k$ and $\{A_j\}_{j=1}^m$ for \emph{I} and \emph{Q} branches, respectively. Fig.~\ref{fig: 4}, illustrates the sub-channel gain profile for all 7 sub-chanels of 128-QAM, before and after stretching the constellation towards the \emph{I}. 

To uniquely recover each sub-channel's data under the presence of the other co-existing interference, it is necessary to have the following relationship between these gains. Condition~(\ref{equ: formation}) is obtained from the layerd formation of constellation.
\begin{equation}
\begin{matrix}
A_p >  \sum_{j=1}^{p-1}d_j.A_j,~~\forall~p > 1\\
B_p >  \sum_{j=1}^{p-1}d_{j+m}.B_j~~~~~~~~~
\end{matrix}
\label{equ: formation}
\end{equation}
where $d_j=\pm1$, and $d_j = 2z_j-1$ (for $z_j$ see Fig.~\ref{fig: 1}). 
\section{Successive Interference Cancellation for detection of $2^n$ QAM}
In this section, successive interference cancellation (SIC) algorithm for detection of $2^n$ QAM in AWGN channels is formulated and discussed. Later, we will show that this algorithm has the same performance as the optimal detection algorithm with table lookup, while the proposed approach has much smaller implementation complexity. In this section we will also show that this algorithm is equivalent to one successive approximation analog to digital conversion (SAR-ADC) in each \emph{I-Q} branch of the baseband receiver.
  
First, let us model the sent and received signal to and from AWGN channel. Based on the vectored implementation model, the baseband transmitter's output $\vec{x}=x_I+jx_Q$ is a weighted sum of bipolar binary data ($\pm 1$) with a number of known gains of $\{B_j\}_{j=1}^k$ and $\{A_j\}_{j=1}^m$. As such, the \emph{I} and \emph{Q} components of ${\vec{x}}$ are:
\begin{equation}
\begin{matrix}
x_Q = d_1.A_1 + d_2.A_2 + \dots + d_m.A_m~~~~~~~~~\\
x_I = d_{m+1}.B_1 + d_{m+2}.B_2 + \dots + d_{m+k}.B_k
\end{matrix}
\label{equ: vectored_Tx}
\end{equation}
In this equations, $d_j = \pm1$, where $d_j=2z_j-1$. $x_I$ and $x_Q$ are respectively $2^k$ and $2^m$ pulse amplitude modulations (PAM) with equally likely discrete and symmetric levels around zero. In baseband AWGN channel, zero mean complex Gaussian noise $\vec{w}$ with mean power of $\sigma^2$ is added to $\vec{x}$, and the received signal at receiver is:
\begin{equation}
\vec{y} = \vec{x} + \vec{w}
\end{equation}
where the \emph{I} and the \emph{Q} components of the received complex signal $\vec{y}$ are:
\begin{equation}
\begin{matrix}
y_Q = d_1.A_1 + d_2.A_2 + \dots + d_m.A_m + w_Q~~~~~~~~~\\
y_I = d_{m+1}.B_1 + d_{m+2}.B_2 + \dots + d_{m+k}.B_k+ w_I
\end{matrix}
\label{equ: vectored_Rx}
\end{equation}
In equation (\ref{equ: vectored_Rx}), $w_I$ and $w_Q$ are zero mean white Gaussian noise with standard deviation of ${\sigma_0}/{\sqrt{2}}$. Now, let us re-arranged (\ref{equ: vectored_Rx}) in the form of:
\begin{equation}
\begin{matrix}
y_Q = d_m.A_m + \sum_{j = 1}^{m-1} d_j . A_j + w_Q~~~~\\
y_I = d_{m+k}.B_k + \sum_{j = 1}^{k-1} d_{m+j} . B_j + w_I
\end{matrix}
\label{equ: vectored_Rx_2}
\end{equation}
In equations (\ref{equ: vectored_Rx_2}), $\sum_{j=1}^{k-1}{d_{m+j}.B_j}$ and $\sum_{j=1}^{m-1}{d_j.A_j}$ are the total interference from the other co-existing sub-channels on the strongest BPSK sub-channel, that according to equation (\ref{equ: formation}) are smaller than $B_k$ and $A_m$, respectively.\\
The objective is to minimize the probability of detection error $P_e$, given the criteria of equation (\ref{equ: formation}).
\begin{equation}
\begin{matrix}
\hat{d}_m = argMin\{P_e(y_Q| A_m > \sum_{j = 1}^{m-1} d_j . A_j)\}~~~~\\
\hat{d}_{m+k} = argMin\{P_e(y_I| B_k > \sum_{j = 1}^{k-1} d_{m+j} . B_j)\}
\end{matrix}
\label{equ: error_optimization}
\end{equation}
The $y_I$ and $y_Q$ are noisy, zero mean PAMs. Apparently, a threshold-based detector with threshold at zero (sign detector) minimizes the error probabilities of equations (\ref{equ: error_optimization}), where it is the maximum likelihood detector for the strongest BPSK sub-channels in \emph{I} and \emph{Q} branches \cite{Gallager}.  
\begin{equation}
\begin{matrix}
\hat{d}_m = Sign(y_Q)\\
\hat{d}_{m+k} = Sign(y_I)
\end{matrix}
\end{equation}
Because $\hat{d}_{m+k}$ and $\hat{d}_m$ are optimum detection for $d_{m+k}$ and $d_m$, then $\hat{d}_{m+k}.B_k$ and $\hat{d}_m.A_m$ are the optimum interference estimates from the the strongest sub-channels that must be cancelled for the next detection step. For this purpose, equations (\ref{equ: vectored_Rx_2}) are re-arranged to the following form:
\begin{equation}
\begin{matrix}
y_Q - \hat{d}_{m}.A_m = d_{m-1}.A_{m-1} + \sum_{j = 1}^{m-2} d_j.A_j + w_Q~~~~~~\\
y_I - \hat{d}_{m+k}.B_k = d_{m+k-1}.B_{k-1} + \sum_{j = 1}^{k-2} d_{m+j}.B_j+ w_I
\end{matrix}
\label{equ: vectored_Rx_3}
\end{equation}
Having cancelled the best estimate of the strongest interference, the sign detector is applied for the best estimate of $\hat{d}_{m+k-1}$ and $\hat{d}_{m-1}$. The process of detection of bits using sign detector continues successively for detection of the other sub-channel's bits in \emph{I-Q} branches, according to (\ref{equ: sign_detector_2}) and (\ref{equ: sign_detector_3}). 
\begin{equation}
\begin{matrix}
\hat{d}_{m-j} = Sign(y_Q-\sum_{p=m-j+1}^{m}\hat{d}_p.A_p)\\
~~~~~~~~~~~~~~~~~~~j = 1, 2, \dots, m-1
\end{matrix}
\label{equ: sign_detector_2}
\end{equation}
\begin{equation}
\begin{matrix}
\hat{d}_{m+k-j} = Sign(y_I-\sum_{p=k-j+1}^{k}\hat{d}_{m+p}.B_p)\\
~~~~~~~~~~~~~~~~~~j = 1, 2, \dots, k-1
\end{matrix}
\label{equ: sign_detector_3}
\end{equation}
This detection algorithm was exploratory introduced in \cite{vectored_QAM} for $2^{2m}$ QAM, based on reverse process of layered formation of the constellation in hierarchical $2^{2m}$ QAM. With this reasoning, the detection algorithm for $2^{k+m}$ QAM with $k$ bits in \emph{I} branch and $m$ bits in \emph{Q} branch is presented in Algorithm-1.

\begin{figure}[b!]
	\centering
		\includegraphics[width=3in,angle=0]{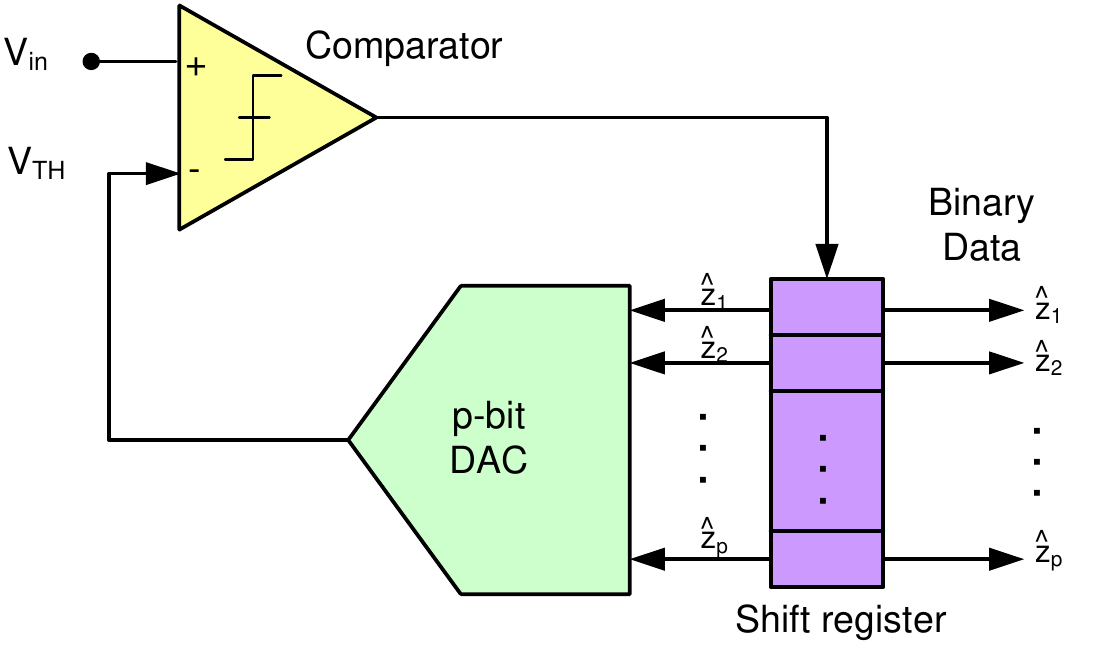}
		\caption{$p$-bit SAR-ADC.}\label{fig: 5}
\end{figure}

\begin{table}[t]
\noindent{Algorithm 1: $2^{n}$-QAM  complexity reduced BSIC detector, $n = k+m$}\\
\line(1,0){250}
\begin{enumerate}
	\item $\hat{\textbf{d}}$ = ($\hat{d}_{m+k}~\cdots~\hat{d}_2~\hat{d}_1$)
	\item $\hat{\textbf{z}}$ = ($\hat{z}_{m+k}~\cdots~\hat{z}_2~\hat{z}_1$)
	\item for $~~p$ = $k: 1: p--$
	\item\quad	 $\hat{d}_{p+m}$ = $Sign(y_I)$,
	\item\quad	 $\hat{z}_{p+m}$ = $[1+Sign(y_I)]/2$,
	\item\quad   $y_I$ = $y_I - \hat{d}_{p+m}.B_p$,
	\item end.	
	\item for $~~p$ = $m: 1: p--$
	\item\quad	 $\hat{d}_p$ = $Sign(y_Q)$,
	\item\quad	 $\hat{z}_p$ = $[1+Sign(y_Q)]/2$,
	\item\quad   $y_Q$ = $y_Q - \hat{d}_p.A_p$,
	\item end.		
\end{enumerate}
\line(1,0){250}
\end{table}
Instead of subtracting and comparing with zero level threshold, we may successively update the detection threshold and then compare $y_I$ and $y_Q$ with the updated thresholds for detection of the next bits. This modification turns the detection algorithm into a SAR-ADC. In $p$-bit SAR-ADC, a given signal in $p$ successive steps is compared with updated thresholds to generate the $p$-bit equivalent digital codeword. This mechanism is illustrated in Fig. \ref{fig: 5}. 

In detection of the strongest sub-channel's bits $\hat{d}_m$ and $\hat{d}_{k+m}$, the SAR-ADC threshold, $V_{TH}$, is zero (see Fig.~\ref{fig: 5}). For the next bits of the \emph{I}-branch and the \emph{Q}-branch, the ADC's thresholds are updated according to $V_{TH,Q,m-j} = \sum_{p=m-j+1}^m \hat{d}_p.A_p$ and $V_{TH,I,m+k-j} = \sum_{p=k-j+1}^{k} \hat{d}_{m+p}.B_p$, respectively. These threshold levels are the summation terms in equations (\ref{equ: sign_detector_2}) and (\ref{equ: sign_detector_3}). \\
\begin{figure}[t]
	\centering
		\includegraphics[width=3in,angle=0]{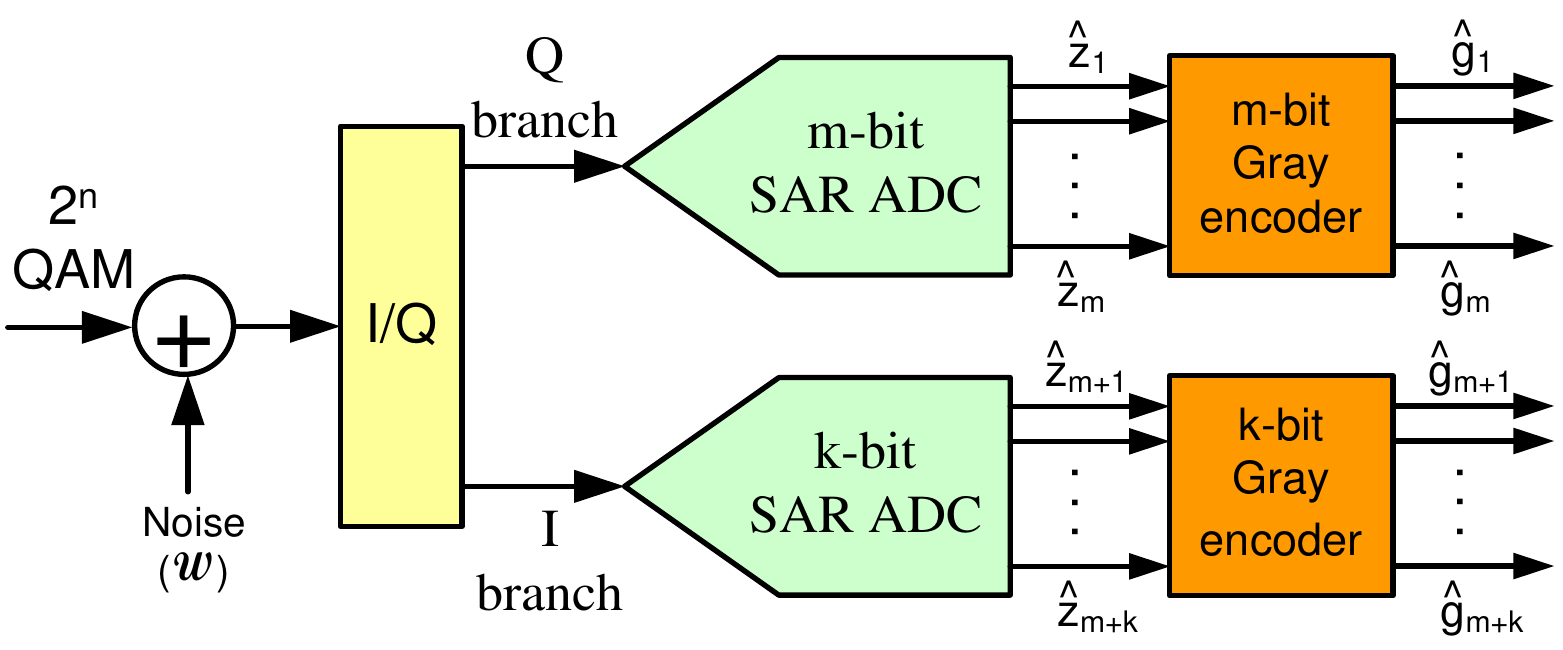}
		\caption{$2^n$ QAM receiver with SAR-ADC in \emph{I} and \emph{Q} branches.}\label{fig: 6}
\end{figure}
A $k$-bit Gray-encoder is an array of $k-1$ exclusive-or (XOR) operations, which is detailed in equations (\ref{equ: Gray_to_Binary}). In this equation, $\hat{z}_k = [1+sign(\hat{d}_k)]/2$, that is $0$ or $1$. 
\begin{equation}
\left\{
\begin{array}{rlrr}
	\hat{g}_k =& \hat{z}_k, \\
	\hat{g}_j=& \hat{z}_j \oplus \hat{z}_{j+1}, & \quad j = 1,\cdots, k-1.
\end{array}
\right.
\label{equ: Gray_to_Binary}
\end{equation}
XOR is a deterministic, memoryless operation. Only one erroneous detection at input to the XOR, results in erroneous output. However, if both of the XOR's input are erroneous, the XOR's output will be correct. Then, the error rate at the output of the Gray encoder will not be more than the error rate of its input. Accordingly, reducing the error rate of its inputs by optimizing the detected bits ($\hat{z}_j$ and $\hat{z}_{j+1}$) at the input of the Gray-encoder, results in optimizing the error rate at the output of the Gray encoder. These SAR-ADCs perform SIC detection. This concludes the following proposition.\\
\textit{Proposition}: In \emph{I} and \emph{Q} branches of baseband $2^{k+m}-QAM$ receiver, respectively $k$ and $m$ bits SAR-ADCs, followed by Gray encoders of the same word size are employed for optimum detection. This is illustrated in Fig.~\ref{fig: 6}. 
\section{Performance Evaluation and Discussion}
In this section, the BER performance of the proposed SIC detection algorithm is presented and it is compared with that of the conventional optimum detector, for $2^n$ QAM. 
First, the BER of 128-QAM with stretched constellation, in AWGN channel once SIC detector is used at receiver is evaluated, using computer simulations. These results then is compared with the performance of optimal QAM detector as it is discussed in \cite{Alouini}.

\begin{figure}
  \centering
  \begin{tabular}{p{.45\columnwidth}p{.45\columnwidth}}
    \includegraphics[width=1.3in,angle=0]{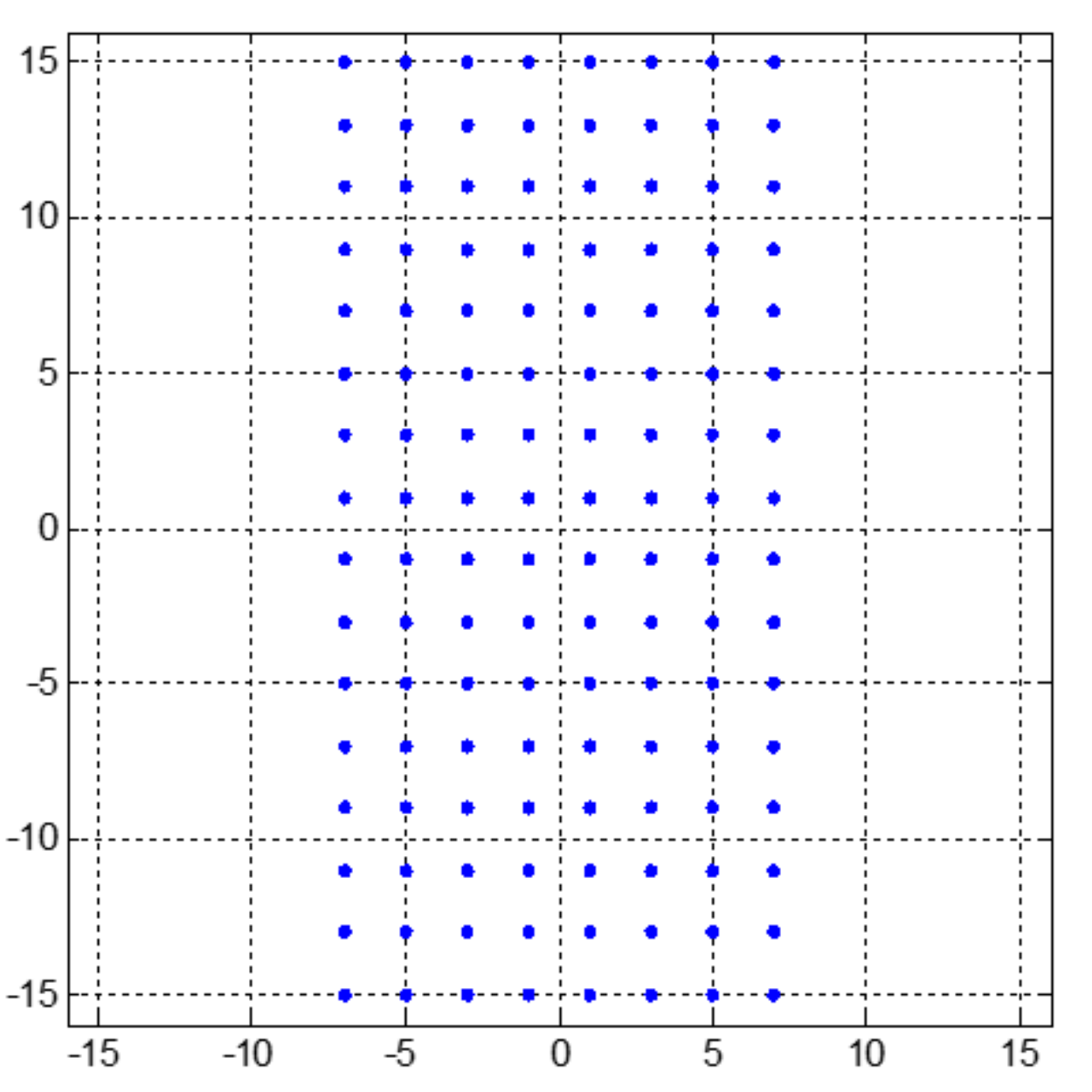}&
    \includegraphics[width=1.3in,angle=0]{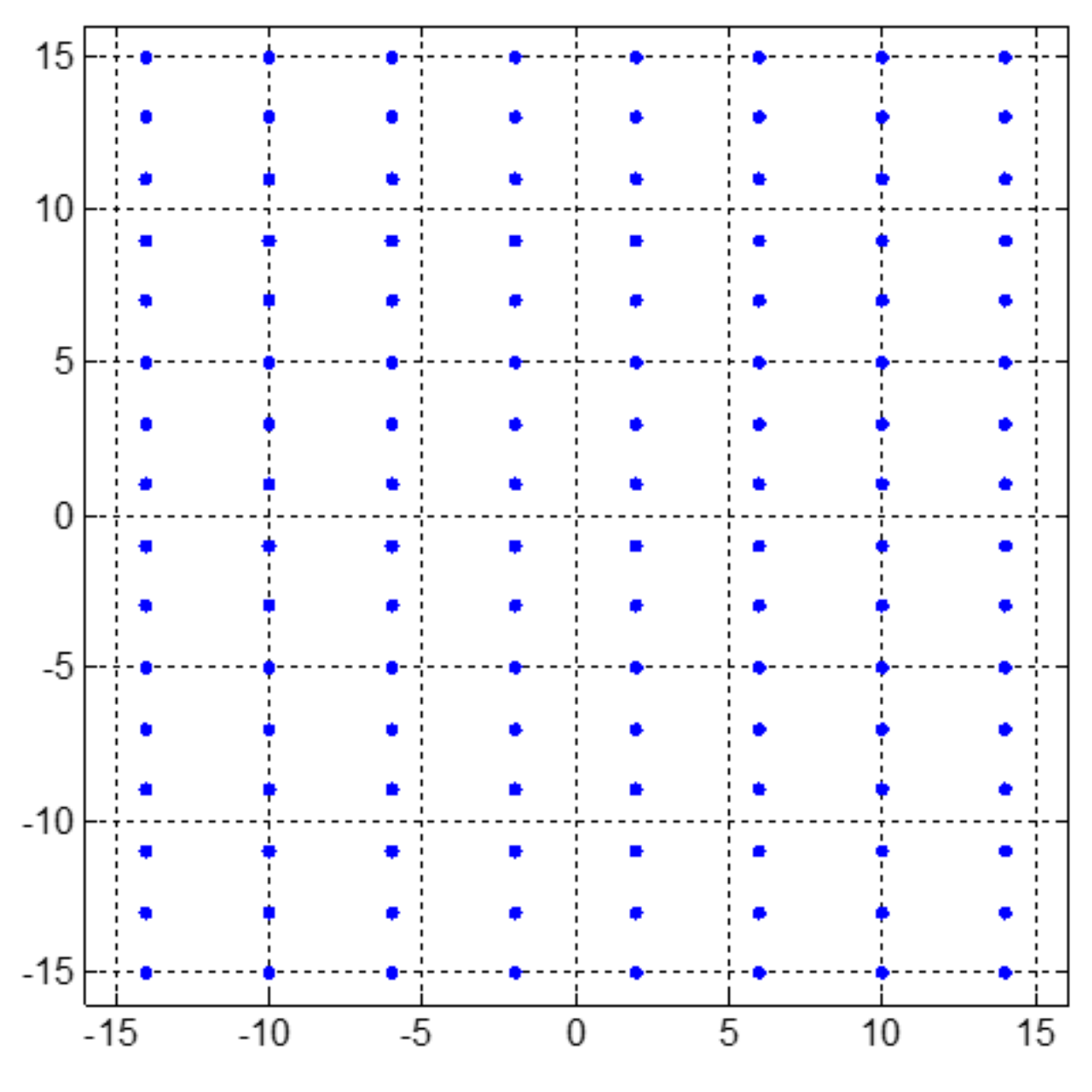}\\
    \footnotesize (a) 128-QAM before stretch. & \footnotesize (b) 128-QAM after stretch ($r$=2).\\
  \end{tabular}  
  \label{fig: 7}\caption{128-QAM before and after constellation stretch with $r$=2.}
\end{figure}

\begin{figure}[t]
   \centering
   \includegraphics[width=3.2in,angle=0]{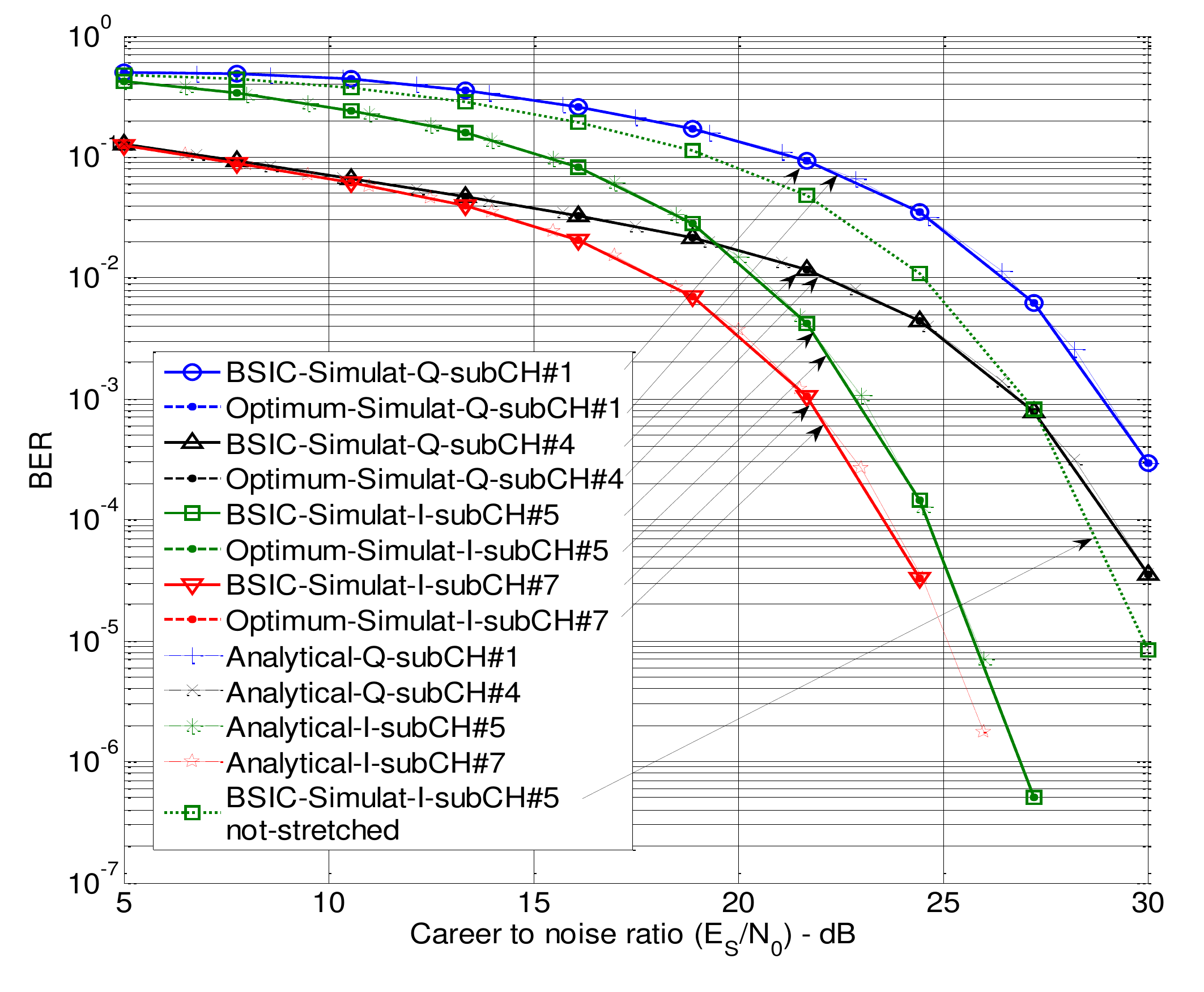}
   \caption{BER comparison of the proposed optimal receiver and the conventional optimal receiver - simulations and analytical}
   \label{fig: 8}
\end{figure}

Fig.~\ref{fig: 8} shows the performance evaluation results. The performance results belongs to the best and the worst sub-channels in \emph{I} and \emph{Q} branches. According to this figure shows, the performance of the proposed approach is the same as that of the optimal receiver, where it is perfectly supported by analytical results. 
As Fig.~\ref{fig: 8} shows, the performance of the \emph{I} bits is considerably better than that of the \emph{Q} bits, which is due to i) smaller number of co-existing interferers on \emph{I} branch (3 bits against 4 bits in \emph(Q) branch); and ii) stretching the constellation towards \emph{I} axis, where it is equivalent to amplifying the sub-channel gain profile for the \emph{I} bits. The sub-channel gain profile is illustrated in Fig.~\ref{fig: 4}. As this Fig.~\ref{fig: 7} shows, the performance of stretched sub-channel $\#5$ at BER of $10^{-5}$ roughly has 4.5 dB over the same not-stretched sub-channel.

In comparing the required storage and complexity, the proposed approach saves a huge amount of storage in comparison to table lookup, which is used in conventional optimum detection, especially for large constellations such as 1024-QAM~\cite{IEEE802.11ax} and 4096-QAM~\cite{Fast-optical, Gorshe}. Table I, presents a brief complexity comparison.\\

For continuing this research, other form of ADCs such as fast sigma-delta ($\Sigma\Delta$) will be investigated. $\Sigma\Delta$ ADC might reduce the effect of additive noise by integration mechanism on its oversampled inputs. This mechanism performs similar to averaging combiner with antenna diversity.

\begin{table}
\caption{Complexity of the proposed versus table lookup, $Max=max(m,k), n=k+m$}\label{Tab: 1}
\begin{tabular}{l||l|l|l}
 \hline
 \hline
  Approach & Computation Load & Required Storage & Gray Code\\
 \hline
 \hline
 Proposed & \begin{tabular}[c]{@{}l@{}l@{}}$Max-1$\\real subtractors\\$Max$ comparators\end{tabular} &                
 \begin{tabular}[c]{@{}l@{}}$n$ storages for  \\ $n$ real values\end{tabular} &   $Max$-bit \\
 \hline
 \begin{tabular}[c]{@{}l@{}}Table \\ Lookup\end{tabular} & No processing load & \begin{tabular}[c]{@{}l@{}}$2^{n+1}$ real  \\  value storages\end{tabular} & $n$-bit \\
 \hline
\end{tabular}
\end{table}%
\section{Conclusion}
A complexity reduced and scalable optimal, baseband detection algorithm is proposed and discussed for hierarchical Gray-coded $2^n$ QAM. The process steps of the optimization algorithm is detailed. The performance of the proposed approach for selected bits are compared with the bit error rate (BER) of the conventional maximum likelihood optimum detection algorithm using Monte-Carlo simulations and analytical approach. The performance evaluation results show a perfect match between the 3 approaches. It is shown that the optimum detector is a successive approximation analog to digital converter in each \emph{I-Q} branch that is followed by a Gray-encoder of the same word size.

\end{document}